\documentclass[12pt]{scrartcl}
\usepackage{amssymb,amsfonts,amsthm,epsfig}

\newtheorem{thm}{Theorem}
\newtheorem{lem}[thm]{Lemma}
\newtheorem{prop}[thm]{Proposition}

\newtheorem*{conH}{Assumption H$_0^m$}
\newtheorem*{conh}{Assumption h$^m$}

\theoremstyle{definition}

\newtheorem{rem}[thm]{Remark}

\newcommand{\er}{\hfill$\diamondsuit$}

\begin{document}

\def\R{{\mathbb R}}
\def\N{{\mathbb N}}
\def\C{{\mathbb C}}
\def\epsi{\varepsilon}

\title{\LARGE\bf
                    Effective $N$-body dynamics for the massless  Nelson model
                    and adiabatic decoupling without spectral gap
}

\author{
\large              Stefan Teufel \\
\normalsize \em     Zentrum Mathematik, Technische Universit\"at M\"unchen,\\[-2mm]
\normalsize \em     80290 M\"unchen, Germany\\[-2mm]
\normalsize         email: teufel@ma.tum.de
}

\date{\normalsize March 22, 2002}

\maketitle

\begin{abstract}
The  Schr\"odinger equation for $N$ particles interacting through effective
pair potentials is derived from  the massless Nelson model with  ultraviolet cutoffs.
We consider a scaling limit where the particles are slow and heavy, but, in contrast to
earlier work \cite{Davies}, no ``weak coupling'' is assumed.
To this end we prove a space-adiabatic theorem without gap condition which gives,
in particular,  control on the rate of convergence in the adiabatic limit.
\end{abstract}

\section{Introduction}

The physical picture  underlying nonrelativistic quantum electrodynamics is that
of charged particles which interact through the exchange of
photons and dissipate energy through emission of photons. In
situations where the velocities of the particles are small
compared to the propagation speed of the photons the interaction
is given through  effective, instantaneous pair potentials.
If, in addition, also accelerations are small, then dissipation through
radiation can be neglected in good approximation.
Instead of full nonrelativistic QED we consider the massless Nelson model. 
This model describes $N$ {\em spinless} particles coupled to a {\em scalar} 
Bose field of zero mass.

The content of this work is a mathematical derivation of the
time-dependent Schr\"o\-din\-ger equation for $N$ particles with Coulombic pair
potentials from the massless Nelson model with ultraviolet cutoffs.
The key mechanism in our derivation is adiabatic decoupling {\em without} a spectral gap.

Before we turn to a more careful discussion of the type of scaling we shall
consider, notice that the coupling of $N$ {\em
noninteracting} particles to the radiation field has {\em
three} effects.
\begin{itemize}
\item
The effective mass, or more precisely, the effective
dispersion relation of the particles is modified. The term
``effective'' refers to the reaction of the particles to weak
external forces. The physical picture is that each particle now
carries a cloud of photons with it, which makes it heavier.
\item
The particles feel an interaction mediated through the field.
 If the propagation speed of the particles is
small compared to the one of the photons, then retardation effects
should be negligible and the interaction between the particles can
be described in good approximation by instantaneous pair
potentials.
\item
Energy is dissipated through photons moving freely to infinity.
The motion of the particles is, in general, no longer of Hamiltonian type.
The rate of energy emitted as photons is proportional to the acceleration of a
particle squared.
\end{itemize}
The scaling to be studied is most conveniently explained on the classical level.
The classical equations of motion for $N$ particles with positions $q_j$, masses $m_j$ and
rigid ``charge'' distributions $\rho_j$  coupled to the scalar field $\phi(x,t)$
with propagation speed  $c$ are
\begin{eqnarray}\label{clf}
\frac{1}{c^2}\,\ddot \phi(x,t)& =& \Delta_x\phi(x,t) - \sum_{j=1}^N\, \rho_j(x-q_j(t)) \\
m_j\,\ddot q_j(t)& =& - \int_{\R^3} dx\,(\nabla_x \phi)(x,t)\,\rho_j(x-q_j(t))
\,,\quad 1\leq j\leq N \,.\label{clp}
\end{eqnarray}
One can think of $\rho_j(x) = e_j\,\varphi(x)$ as a smeared out
point charge $e_j$ with a form factor $\varphi\geq 0$ satisfying $\int_{\R^3} dx\,\varphi(x)=1$.
Taking the limit $c\to \infty$
in (\ref{clf}) yields the Poisson equation for the field and thus, after elimination of
the field, (\ref{clp}) describes $N$ particles interacting through smeared
Coulomb potentials. Mass renormalization for the particles in not visible at leading
order.

Instead of taking $c\to \infty$ one can as well explore for which scaling of
the particle properties one obtains analogous effective equations.
Since retardation effects should be negligible, the initial velocities of the
particles are now assumed to be $\mathcal{O}(\epsi)$ compared to the fixed propagation
speed $c=1$ of the field, $\epsi\ll 1$.
In order to see motion of the particles over finite distances, we
have to follow this dynamics at least over times of order $\mathcal{O}(\epsi^{-1})$.
To make sure that the velocities are still of order $\mathcal{O}(\epsi)$ after times
of order $\mathcal{O}(\epsi^{-1})$, the accelerations must be at most of order
$\mathcal{O}(\epsi^2)$. The last constraint also guarantees that the energy
dissipated over times of order $\mathcal{O}(\epsi^{-1})$ is at most of
order $\mathcal{O}(\epsi^3)$.

The  natural procedure would now be to consider such initial data, for which the
velocities stay of order $\mathcal{O}(\epsi)$ over sufficiently long times. 
The problem simplifies if we assume, as we shall do in this work, 
that the mass of a particle is of order $\mathcal{O}(\epsi^{-2})$. As a consequence
accelerations are and stay of order $\mathcal{O}(\epsi^2)$ {\em uniformly for
all initial conditions}. In this scaling limit mass renormalization is {\em not}
visible at leading order. Indeed, if we substitute $t'=\epsi t$
and $m_j'=\epsi^2 m_j$ in (\ref{clf}) and (\ref{clp}), we find
that the limit $\epsi\to 0$ is equivalent to the limit $c\to\infty$.

After quantization, however, the two limiting procedures  are no longer
equivalent. The limit $c\to \infty$ for the
Nelson model was analyzed by Davies \cite{Davies} and later also by Hiroshima
\cite{Fumio}, who removed the ultraviolet cutoff. A comparison of their results
with ours can be found at the end of this introduction.
We will adopt the point of view that it is more natural to
explore the regime of particle properties which gives rise to effective
equations than to take the limit $c\to\infty$.

The deeper reason for our choice is that the more natural procedure 
of restricting to appropriate initial conditions gives rise to
a similar mathematical structure. If the bare mass of the particles 
of order $\mathcal{O}(1)$, then the proper scaling which
yields effective equations with renormalized masses was introduced
and analyzed  for the classical Abraham model by Kunze and
Spohn, see \cite{KS1,Spohn} and references therein. 
Denoting again the ratio of the velocities of the particles and the field
as $\epsi$, they consider
charges initially separated by distances of order $\mathcal{O}(\epsi^{-2})$
in units of their diameter. Hence 
the forces are $\mathcal{O}(\epsi^4)$ initially. 
For times up to order $\mathcal{O}(\epsi^{-3})$
and for appropriate initial conditions -- excluding head on collisions --
the separation of the particles remains of order $\mathcal{O}(\epsi^{-2})$ and thus the
velocities remain of order $\mathcal{O}(\epsi)$.
In particular, the rescaled macroscopic position $q'(t') = \epsi^{-2}\, q(t'/\epsi^3)$
satisfies $(d/d t')^2 \, q'(t') = \mathcal{O}(\epsi^4)$, which matches
the order of the forces. As a consequence 
one obtains a sensible limiting dynamics for the macroscopic variables.

One would expect that the same scaling limit applied to the
quantum mechanical model yields in a similar fashion effective dynamics with 
renormalized dispersion. However, inserting this scaling into the massless 
Nelson model, one faces  mathematical  problems beyond those in the 
simpler  $m=\mathcal{O}(\epsi^{-2})$ scaling. Without going into details
we remark that the main problem is that for massless bosons the Hamiltonian 
at fixed total momentum does not have a ground state in Fock space, cf.\ \cite{Froe,Chen}.
(As a consequence it is not even  clear how to translate
the result in \cite{TS} for a {\em single quantum} particle coupled to
a {\em massive}  quantized scalar field and subject to weak external forces
to the  massless case.)
Nevertheless, the simpler scaling with $m=\mathcal{O}(\epsi^{-2})$ provides at least
a first step in the right direction, since the mechanism of adiabatic decoupling without
gap will certainly play a crucial role also in a more refined analysis.

In the remainder of the introduction we briefly present the massless Nelson model,
explain our main result and compare it to Davies' ``weak coupling limit'' \cite{Davies}.

Up to a modified dispersion for the particles, the following model is obtained
through canonical quantization of the classical system (\ref{clf}) and (\ref{clp}).
The state space for $N$ spinless particles is $L^2(\R^{3N})$
and as Hamiltonian we take
\begin{equation} \label{Hp1def}
H_{\rm p} = \sum_{j=1}^N \, \sqrt{-c_{\rm max}^2\Delta_{x_j} + c_{\rm max}^4m^2}\,,
\end{equation}
where $c_{\rm max}$ is the maximally attainable speed of the particles and $m$ their mass, $\hbar=1$.
As explained before, we consider the scaling limit
\begin{equation} \label{scaling}
\epsi\ll 1 \quad{\rm with}\quad c_{\rm max}
=\mathcal{O}(\epsi)\quad{\rm and} \quad m=\mathcal{O}(\epsi^{-2})\,.
\end{equation}
It might seem somewhat artificial to have a relativistic
dispersion relation for the particles which does not contain the
speed of light, but some other maximal speed $c_{\rm max}$. This is done
only for the sake of simple presentation.
We could as well consider the quadratic dispersion
$H_{\rm p} = - \sum_{j=1}^N\, \frac{1}{2m} \Delta_{x_j}$ for the
particles. However, there would be no maximal speed and we would be forced
to either introduce a cutoff for large momenta or to change the topology in
(\ref{result}). While both strategies are
technically straightforward by using exactly the same methods
as in \cite{ST} in the context of Born-Oppenheimer approximation,
they would obscure the simple structure of our result.

We insert the scaling (\ref{scaling}) into (\ref{Hp1def}) and
change units such that the particle Hamiltonian is now given through
\begin{equation}\label{Hpepsi}
H^\epsi_{\rm p} = \sum_{j=1}^N \, \sqrt{-\epsi^2\Delta_{x_j} + 1}\,.
\end{equation}
The particles are coupled to a scalar field whose state is an element of the bosonic
Fock space over $L^2(\R^3)$ given as
\begin{equation}\label{Fock}
\mathcal{F} = \oplus_{m=0}^\infty \otimes^m_{\rm (s)} L^2(\R^3)\,,
\end{equation}
where $\otimes^m_{\rm (s)}$ is the $m$-times symmetric tensor product and
$\otimes^0_{\rm (s)} L^2(\R^3) :=\C$.
The  Hamiltonian for the free bosonic field is
\begin{equation}\label{Hf}
H_{\rm f} = {\rm d}\Gamma(|k|)\,,
\end{equation}
where $k$ is  the boson momentum.
In our units  the propagation speed of the bosons is equal to one.
The reader who is not familiar with the notation is asked to consult the
beginning of Section 3, where the model is introduced in full detail.

In the standard Nelson model the coupling between the $j^{th}$ 
particle and  the field
is given through
\begin{equation} \label{fieldop}
H_{{\rm I}, j} = \int_{\R^3} dy\,\phi(y)\,\rho_j(y-x_j)\,,
\end{equation}
where $\phi$ is the field operator in position representation and $x_j$ the position
of the $j^{th}$ particle. The charge density  $\rho_j\in L^1(\R^3)\cap L^2(\R^3)$
of the $j^{th}$ particle is
assumed to be spherically symmetric and
its Fourier transform is denoted by  $\hat \rho_j$. For the moment we also assume
an infrared condition, namely that
\begin{equation}\label{IRCon}
\sum_{j=1}^{N}\frac{\hat\rho_j(k)}{|k|^{3/2}} \in L^2(\R^{3})\,.
\end{equation}
Condition (\ref{IRCon}) constrains the  total charge
of the system but {\em not} that of an individual particle to zero.
The state of the combined particles + field system is an element of
\[
\mathcal{H} = L^2(\R^{3N})\otimes \mathcal{F}
\]
and its time evolution is generated
by the Hamiltonian
\begin{equation} \label{HepsiDef}
H^\epsi = H^\epsi_{\rm p}\otimes{\bf 1} + {\bf 1}\otimes {\rm d}\Gamma(|k|)
+ \sum_{j=1}^N H_{{\rm I},j}\,.
\end{equation}
Note that $H$ contains no terms which directly couple different particles. All interactions
between the particles must be mediated through the boson field.

Our goal is the  construction of approximate solutions of the time dependent Schr\"odin\-ger
equation
\begin{equation} \label{Sfull}
i\epsi\frac{d}{dt}\Psi(t) = H^\epsi \Psi(t)\,,\quad \Psi(0)=\Psi_0\in \mathcal{H}
\end{equation}
from solutions of an effective Schr\"odinger equation
\begin{equation}\label{Seff}
i\epsi\frac{d}{dt}\psi(t) = H^\epsi_{\rm eff} \psi(t)\,,\quad \psi(0)=\psi_0\in L^2(\R^{3N})
\end{equation}
for the particles only. Notice the factor $\epsi$ in front of the time derivative in
(\ref{Sfull}) and (\ref{Seff}), which means that we switched to a time scale of order
$\epsi^{-1}$ in microscopic units. As explained before,
this is necessary in order to see nontrivial dynamics
of the particles, since their speed is $\mathcal{O}(\epsi)$.

We remark that the scaling (\ref{scaling}) coincides with the one in time-dependent
 Born-Oppenheimer approximation, where $m=\mathcal{O}(\epsi^{-2})$ is the
mass of the nuclei and where, at fixed kinetic energy, the
velocities of the nuclei are also of order $\mathcal{O}(\epsi)$.
The Hamiltonian (\ref{HepsiDef})
has the same structure as the molecular Hamiltonian and the role of the electrons
in the Born-Oppenheimer approximation is now played by the bosons.
\begin{figure}
\begin{center}
\epsfig{file=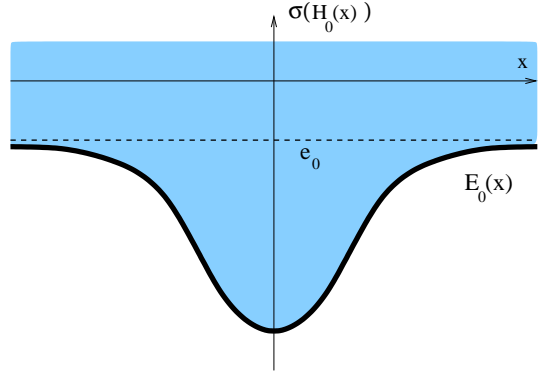, height=5cm, scale=1}
\end{center}
\caption{The spectrum of $H_0(x)$ for $N=2$. The thick line indicates the eigenvalue
$E_0(x)$ sitting at the bottom of continuous spectrum. }
\end{figure}

The key observation for the following is that the interaction Hamiltonian depends
only on the configuration $x$ of the particles and that the operator
\[
H_0(x) =  {\rm d}\Gamma(|k|) +
\sum_{j=1}^N H_{{\rm I},j}(x)\,,
\]
which acts on $\mathcal{F}$ for fixed $x\in \R^{3N}$, has a unique ground state
$\Omega(x)$ with ground state energy
\begin{equation}\label{E0Def}
E_{0}(x) =\sum_{j=2}^N\sum_{i=1}^{j-1} V_{ij}(x_i - x_j) + e_0\,,
\end{equation}
 where
\begin{equation}\label{vijDef}
V_{ij}(z) = - \int_{\R^3\times\R^3}dv\,dw\, \frac{\rho_i(v-z)\rho_j(w)}{4\pi|v-w|}
\end{equation}
and
\begin{equation}\label{e0Def}
 e_0 = -
\frac{1}{2}\sum_{j=1}^N\int_{\R^3\times\R^3}dv\,dw\,\frac{\rho_j(v)\rho_j(w)}{4\pi|v-w|}\,.
\end{equation}
$V_{ij}(z)$ is the electrostatic interaction energy of the charge distributions
$\rho_i$ and $\rho_j$ at distance $z$, however, with the ``wrong'' sign. It is a peculiarity
of the scalar field that the interaction between charges with equal sign is attractive.
$e_0$ is the sum of all self energies.
The remainder of the spectrum is purely absolutely continuous and the ground state energy
is {\em not} isolated, cf.\ Figure 1.

Let $P_*(x) = |\Omega(x)\rangle\langle \Omega(x)|$, then the states in
\begin{equation}\label{RanP}
{\rm Ran}\,P_* = \left\{ \int^\oplus_{\R^{3N}}dx\,
\psi(x)\Omega(x):\,\,\psi\in L^2(\R^{3N})
\right\}\subset \mathcal{H}
\end{equation}
correspond to  wave packets without free bosons. If the particles are moving at small
speeds and if the accelerations are also small, one expects that
no free bosons are created, i.e.\ that Ran$P_*$ is approximately invariant
under the dynamics generated by $H^\epsi$. Moreover the wave function $\psi(x)$
of the particles should approximately be governed by the effective Schr\"odinger
equation (\ref{Seff}) with
\[
H^\epsi_{\rm eff} =  \sum_{j=1}^N \, \sqrt{-\epsi^2\Delta_{x_j} + 1} +
 \sum_{j=2}^N\sum_{i=1}^{j-1} V_{ij}(x_i - x_j)\,.
\]
Our main result, Theorem \ref{NelsonTheorem}, states that for
$\Psi_0 =\int^\oplus dx\,\psi_0(x)\Omega(x)\in$ Ran$P_*$ we have
\begin{equation} \label{result}
\Big\| e^{-iH^\epsi t/\epsi} \Psi_0 -
\int^\oplus_{\R^{3N}} dx
\left(e^{-iH_{\rm eff}^\epsi t/\epsi}\,\psi_0\right)(x)\,e^{-ie_0t/\epsi}\,\Omega(x)
\Big\| = \mathcal{O}(\epsi\sqrt{\ln(1/\epsi)})\,(1+|t|)\,\|\Psi_0\|\,.
\end{equation}
Notice that in the approximate solution of the full Schr\"odinger
equation the state of the field is, up to a fast oscillating
global phase $e^{-ie_0t/\epsi}$, adiabatically following the
motion of the particles. In particular, there are no bosons
traveling back and forth between the particles and the phrase
that the particles ``interact through the exchange of
bosons'', which comes from perturbation theory, should not be
taken literally in the present setting.

As mentioned before, there is a strong similarity to the
time-dependent Born-Oppen\-heimer approximation, where one obtains
an effective Schr\"odinger equation for the nuclei in a molecule
with an effective potential generated by the electrons \cite{ST}.
In both cases the physical mechanism which leads to the
approximate invariance of the subspace Ran$P_*$ is adiabatic
decoupling. I.e.\ the separation of time scales for the motion of
the different parts of the system lets the fast degrees of
freedom, in our case the bosons, instantaneously adjust to the
motion of the slow degrees of freedom, the particles.

However, for massless bosons -- in contrast to the Born-Oppenheimer
approximation -- there is no spectral gap which pointwise separates the
energy band $E_0=\{(x,E_0(x)): x\in \R^{3N}\}$ from
the remainder of the spectrum of $H_0(x)$, but $E_0$ lies at the bottom of continuous spectrum.
Hence we need a space-adiabatic theorem, cf.\ \cite{ST,TS,PST},  {\em without} gap condition.
Only recently time-adiabatic theorems without a gap condition were established in
\cite{AE,Bornemann,Note}. The notion time-adiabatic refers to the
setting where the Hamiltonian of the system is itself time-dependent and varies on a slow
time scale.
In Section 2 a general space-adiabatic theorem without gap condition is formulated and proved.
The proof is based on ideas developed in \cite{Note} and our approach gives, in particular, good
control on the rate of convergence in the adiabatic limit.

As an application of the result from Section 2 we consider in Section 3 the scaling
limit $\epsi\to 0$ of the massless Nelson model as described above. We emphasize at this point
that, in view of the missing gap condition, the rate of convergence
$\mathcal{O}(\epsi\sqrt{\ln(1/\epsi)})$
in (\ref{result})  is surprisingly fast, since it is almost as good as in the case with a gap.
Moreover, if all particles have individually total charge equal to zero, then the rate is
exactly $\mathcal{O}(\epsi)$ as in the case with a gap. Hence, the logarithmic
correction must be attributed to the Coulombic long range character of the interparticle
interaction.
In the situation with gap it is known \cite{MS,PST} that
the wave function stays in a subspace
Ran$P_*^\epsi$ which is $\epsi$-close to the band subspace Ran$P_*$ up to an error of
order $\mathcal{O}(\epsi^\infty)$.
However,  in the situation
without gap, we expect that a piece of order $\epsi^\alpha$, $\alpha<\infty$,
of the wave function
is ``leaking out'' of Ran$P_*$ in the sense that it becomes orthogonal to Ran$P_*$
under time evolution. Physical considerations suggest that
$\alpha=3/2$ for the present problem, see Remark \ref{leakrem}.
As a consequence,  the $\epsi^2$ corrections to the effective Hamiltonian are still
dominating dissipation and can be formally derived from the results in \cite{PST}.
The effective Hamiltonian (\ref{Darwin}) then contains a renormalized mass term
and the momentum dependent Darwin interaction.

Finally let us compare our results to those obtained by Davies \cite{Davies}, who considers
the limit $c\to\infty$ for the Hamiltonian
\begin{equation}
H^c = H_{\rm p} \otimes {\bf 1} + {\bf 1} \otimes {\rm d}\Gamma(c|k|) + \sqrt{c} H_{\rm I}\,.
\end{equation}
Notice that $H^c$ is obtained through canonical quantization of (\ref{clf}) and (\ref{clp})
if one does not set $c=1$ as we did before.
Davies proves that for all $t\in\R$
\[
{\rm s}-\lim_{c\to\infty} e^{-iH^c t}(\psi\otimes\Omega) = (e^{-i(H_{\rm eff}+e_0) t}\psi)\otimes\Omega\,,
\]
where $\Omega=\{1,0,0,\ldots\}$ denotes the Fock vacuum and $H_{\rm p} :=H^{\epsi =1}_{\rm p}$
and $H_{\rm eff}:= H^{\epsi=1}_{\rm eff}$. This shows that although the limit $c\to\infty$
is equivalent to our scaling on the classical level, the results for the quantum model differ
qualitatively. While we obtain effective dynamics for states which contain a nonzero
number of bosons independent of $\epsi$, cf.\ (\ref{result}),
the $c\to \infty$ limit
yields effective dynamics for states which contain no bosons at all.
Furthermore, the limit $\epsi\to 0$ is a singular limit as no limiting dynamics
for $\epsi=0$ exists.

In Section 3 we also consider
an infrared-renormalized model suggested in \cite{Arai} and \cite{LMS},
which allows us to do without the global infrared condition (\ref{IRCon}). The results are
exactly the same as in the standard Nelson model.

\section{A space-adiabatic theorem without gap condition}

Generalizing from the time-adiabatic theorem of quantum mechanics \cite{Kato},
we consider perturbations of self-adjoint operators $H_0$, which are
fibered over the base space $\R^n$, where, for better readability, we use
$M:=\R^n$ to denote this base space.
Let  $\mathcal{H}_{\rm f}$ be a separable Hilbert space and let $dx$ denote
Lebesgue measure.
Recall that $H_0$ acting on $\mathcal H = L^2(M,dx)\otimes \mathcal{H}_{\rm f}
=L^2(M, dx; \mathcal{H}_{\rm f})$
is called fibered, cf.\ \cite{RS4}, if
there is a measurable map $M\ni x\mapsto H_0(x)$ with values in the self-adjoint
operators on $\mathcal{H}_{\rm f}$ such that
\[
H_0 = \int^\oplus_M\,dx\,H_0(x)\,.
\]
There seems to be no standard name for the set
\[
\Sigma = \Big\{(x,s)\in M\times \R, s\in\sigma(H_0(x))\Big\}
\]
and we propose to call it the {\em fibered spectrum} of $H_0$.

Let $\sigma_*\subset\Sigma$ be such that $x\mapsto P_*(x)$ is
measurable, where $P_*(x)$ denotes the
spectral projection of $H_0(x)$ associated with $\sigma_*(x)$.
Then $P_* = \int^\oplus_M\,dx\,P_*(x)$ is an orthogonal
projection which commutes with $H_0$, but which is in general not
a spectral projection of $H_0$.

We  consider perturbations of $H_0$, which mix the fibers, in a sense,
slowly. As a prototype consider for a sufficiently regular real valued function $h$ on
``momentum space'' $\R^n$ the   self-adjoint operator $h^\epsi = h(-i\epsi\nabla_x)$
 on $L^2(M)$. Here $\epsi>0$ is the adiabatic parameter and
$[h^\epsi \otimes {\bf 1}, A] = \mathcal{O}(\epsi)$ for any operator
$A$ which is fibered over $M$. Let
\[
H^\epsi = H_0 + h^\epsi \otimes {\bf 1}\,,
\]
 then the invariant  subspaces for $H_0$ constructed above are still ``approximately''
invariant for $H^\epsi$ with $\epsi$ small,
since $[H^\epsi,P_*]=\mathcal{O}(\epsi)$ and thus
$[e^{-iH^\epsi s},P_*]= \mathcal{O}(\epsi|s|)$. But  the relevant time scale
for the dynamics generated by $h^\epsi$ is ${t}/{\epsi}$ with $t=\mathcal{O}(1)$.
Thus the unitary group of interest is $e^{-iH^\epsi {t}/{\epsi}}$.  However, according to
the naive argument, $[e^{-iH^\epsi {t}/{\epsi}},P_*]= O(|t|)$ and the
subspaces Ran$P_*$ seem to be not even approximately invariant as $\epsi \to 0$.

It is well known \cite{ST,TS} that the failure of the naive
argument can be cured if $\sigma_*$ is separated by a gap from the remainder of the fibered
spectrum $\Sigma$.
Then  $[e^{-iH^\epsi {t}/{\epsi}},P_*]= \mathcal{O}(\epsi)\,(1+|t|)$, a result that
was baptized space-adiabatic theorem in \cite{ST}.
The object of this section is to establish an analogous result without assuming
a gap condition.

We remark that the general setup for space-adiabatic theory are
Hamiltonians which are ``fibered'' over phase space, in the sense
that they can be written as quantizations of operator valued
symbols \cite{PST}.

\subsection{Assumptions and results}

Let $H_0(x)$, $x\in M$, be a family of self-adjoint operators  on some
common dense domain $\mathcal{D}\subset\mathcal{H}_{\rm f}$, $\mathcal{H}_{\rm f}$  a
separable Hilbert space.
Let $\|\cdot\|_{H_0(x)}$ denote the graph norm of $H_0(x)$ on $\mathcal D$, i.e.,
for $\psi\in \mathcal D$, $\|\psi\|_{H_0(x)} = \|H_0(x)\psi\|+\|\psi\|$. We assume
that all the $H_0(x)$-norms
are equivalent in the sense that there is an $x_0\in M$ and  constants
$C_1,C_2 <\infty$ such that
$ C_1 \|\psi\|_{H_0(x_0)}\leq\|\psi\|_{H_0(x)} \leq C_2 \|\psi\|_{H_0(x_0)}$.
Then
\[
H_0 = \int_M^\oplus\,dx\,H_0(x)
\]
 with domain
$D(H_0)=L^2(M)\otimes \mathcal{D}$ is self-adjoint, where here and in the following
$\mathcal D$ resp.\ $D(H_0)$ are understood to be equipped with the $\|\cdot\|_{H_0(x_0)}$
resp.\ $\|\cdot\|_{H_0}$  norm.
For $k\in\N_0$ and $\mathcal{E}$ some Banach space let
\[
C_{\rm b}^k (\R^n,\mathcal{E}) = \left\{ f\in C^k (\R^n,\mathcal{E}):
\sup_{x\in\R^n} \| \partial^\alpha_x f (x) \|_\mathcal{E} < \infty
\quad\forall\,\alpha\in\N^n \quad{\rm with}\quad|\alpha|\leq k \right\}\,.
\]
$\mathcal{L}(\mathcal{H}_1,\mathcal{H}_2)$ denotes the space of bounded linear
operators from $\mathcal{H}_1$ to $\mathcal{H}_{2}$ and $\mathcal{L}_{\rm sa}(\mathcal{H})$ denotes
the set of bounded self-adjoint operators on $\mathcal{H}$. Let
$|\cdot|$ be the Euclidean norm on $\R^n$ and denote
the Hessian of a function $A$ on $\R^n$ by $\nabla^{(2)}A(x)$.
For the resolvent we write $R_\lambda(A) = (A-\lambda)^{-1}$.
Let $m\geq 2$.
\begin{conH}
Let $H_0(\cdot)\in C^{m}_{\rm b}(M, \mathcal{L}(\mathcal{D},\mathcal{H}_{\rm f}))$ and
for all $x\in M$ let $P_*(x)$ be an orthogonal projection such that
$H_0(x)\,P_*(x) = E(x)\,P_*(x)$ with
$P_*(\cdot)\in C^{m+1}_{\rm b}(M, \mathcal{L}(\mathcal{H}_{\rm f}))$
and $E(\cdot) \in C^{m}_{\rm b}(M, \R)$.

In addition one of the following assertions holds:
\begin{enumerate}
\item  For  $1\leq j\leq n$
\begin{equation}\label{NGB0}
\lim_{\delta\to 0}\,
{\rm ess}\sup_{\hspace{-.5cm} x\in M}
 \|\,\delta\,R_{E(x)-i\delta}(H_0(x))\,
(\partial_{x_j} P_*)(x)P_*(x)\|_{\mathcal{L(H}_{\rm f})}=0\,.
\end{equation}
\item
There is a constant $\delta_0>0$ and a function $\eta:[0,\delta_0]\to [0,\delta_0]$
with $\eta(\delta)\geq\delta$
 and a constant  $C<\infty$ such that for $\delta \in (0,\delta_0]$ and  $1\leq j\leq n$
\begin{equation}\label{NGB1}
{\rm ess}\sup_{\hspace{-.5cm} x\in M}
 \|\,R_{E(x)-i\delta}(H_0(x))\,(\partial_{x_j} P_*)(x)P_*(x)\|_{\mathcal{L(H}_{\rm f})}\leq C\,\delta^{-1}\,
\eta(\delta)\,.
\end{equation}
\item
In addition to (\ref{NGB1}) for $1\leq k,j\leq n$ also
\begin{equation}\label{NGB2}
{\rm ess}\sup_{\hspace{-.5cm} x\in M}
\left\| \partial_{x_k} \Big(\, R_{E(x)-i\delta}(H_0(x)) (\partial_{x_j} P_*)(x)P_*(x)\Big)
\right\|_{\mathcal{L(H}_{\rm f})}\leq
C\,\delta^{-1}\,\eta(\delta)
\end{equation}
holds.
\end{enumerate}
\end{conH}
\noindent A few remarks concerning Assumption  {\bf H$^m_0$} are in order:
\begin{itemize}
\item
It is {\em not} assumed that
$P_*(x)$ is the spectral projection of $H_0(x)$ corresponding to the eigenvalue
$E(x)$. However, (\ref{NGB0})  holds pointwise in $x$ whenever
$P_*(x)$ is the spectral projection and has finite rank, cf.\ Proposition \ref{PropAE}.
\item
Inequality (\ref{NGB1}) is always satisfied with $\eta(\delta)=1$.
For Assumption  {\bf H$^m_0$} (ii) and (iii) to have nontrivial consequences
on the rate of convergence in the adiabatic theorem,
$\eta(\delta)$ must satisfy $\lim_{\delta\to 0}\eta(\delta)=0$.
These assumptions might look rather artificial at first sight,
but turn out to be very natural in the proof and also in our
application. For the simpler time-adiabatic setting, which gives
rise to similar conditions, we refer the reader to \cite{Note}.
\item
The regularity of $P_*(x)$ has to be
assumed, since it does not follow from the regularity of $H_0(x)$
without the gap condition, even if $P_*(x)$ is spectral.
The regularity of $E(x)$ follows from the one of
 $H_0(x)$ and $P_*(x)$ whenever $P_*(x)$ has finite rank, as can be seen by writing
$E(x) = \mbox{tr}(H_0(x)P_*(x))/\mbox{tr}P_*(x)$.
\end{itemize}

The ``band subspace'' Ran$P_*$ defined through $P_* =
\int^\oplus_M dx\, P_*(x)$ is invariant under the dynamics
generated by $H_0$, since $[H_0,P_*]=0$ holds by construction.
We will consider  perturbations $h^\epsi$ of $H_0$ satisfying

\begin{conh}
For $\epsi\in (0,1]$ let $h^\epsi$ be a self-adjoint
operator with domain $D(h)\subset\mathcal{H}$ independent of $\epsi$ such that
$H_0 + h^\epsi$ is essentially self-adjoint on $D(h)\cap D(H_0)$.
There exists an operator
$(D h)^\epsi \in \mathcal{L}_{\rm sa}(\mathcal{H})^{\oplus n}$
with $\sup_{\epsi\in(0,1]} \|\,|\,(D h)^\epsi\,|\,\|_{\mathcal{L}(\mathcal{H})}<\infty$
satisfying:
\begin{enumerate}
\item There is a constant $C<\infty$ such that
 for each $A\in C^m_{\rm b}(M,\mathcal{L}(\mathcal{H}_{\rm f}))$
\[
\| [h^\epsi, A] + i\,\epsi \,\nabla_xA\cdot(D h)^\epsi \|_{\mathcal{L}(\mathcal{H})}\leq
 \,C\,\sum_{j=2}^m \,\epsi^j \sup_{x\in M,\,|\alpha|=j} \|\partial_x^\alpha
A(x)\|_{\mathcal{L}(\mathcal{H}_{\rm f})}\,.
\]
\item There is a constant $C<\infty$ such that
\[
 \|\,|\, [(D h)^\epsi, H_0 ]\,|\,\|_{\mathcal{L}(D(H_0),\mathcal{H})}+\|\,|\,[(D h)^\epsi,
 h^\epsi ]\,|\,\|_{\mathcal{L}(\mathcal{H})} \leq \epsi\, C \,.
\]
\end{enumerate}
\end{conh}

By assumption, $H^\epsi = H_0 + h^\epsi$ is essentially self-adjoint on
$D(h)\cap D(H_0)$ and we use its closure, again denoted by $H^\epsi$, to define
for  $t\in\R$
\[
U^\epsi(t) = e^{-i  H^\epsi t/\epsi}\,.
\]
Since, according to Assumption {\bf h}$^m$ (i), 
$[H^\epsi,P_*]=[h^\epsi,P_*]=\mathcal{O}(\epsi)$, the naive
argument gives $[U^\epsi(t),P_*]= |t|\mathcal{O}(1)$. Indeed, our aim is to
cure the failure of the naive argument and to show that
Ran$P_*$ is invariant for $U^\epsi(t)$ in the limit $\epsi\to 0$. To this end
 we will compare $U^\epsi(t)$ with the unitary group generated by
\[
H^\epsi_{\rm diag} \,=\,H_0\,+\, P_*\,h^\epsi
 \,P_* \,+\, P_*^\perp\,h^\epsi \,P_*^\perp\,.
\]
Also $H^\epsi_{\rm diag}$ is self-adjoint on $D(H^\epsi)$
since $P_*(\cdot)\in
C^m_{\rm b}(M,\mathcal{L}(\mathcal{H}_{\rm f}))$
and thus $H^\epsi-H^\epsi_{\rm diag}=P_*^\perp [h^\epsi,P_*]P_* -
P_*[h^\epsi,P_*]P_*^\perp $ is bounded according to {\bf h$^m$} (i).
Again we abbreviate for $t\in\R$
\[
U^\epsi_{\rm diag}(t) = e^{-i H^\epsi_{{\rm diag}} t /\epsi}\,,
\]
and we have by construction that
\[
\big[ P_*, U^\epsi_{\rm diag} (t)] = 0\,,
\]
i.e.\ Ran$P_*$ and Ran$P_*^\perp$ are invariant subspaces for the dynamics generated
by $ H^\epsi_{{\rm diag}}$.

\begin{thm}\label{NoGapTheorem}
Assume {\bf H$_0^m$} and  {\bf h$^m$} for some $m\geq 2$. Let
 $\epsi\in(0,\delta_0]$, then
\begin{itemize}
\item  {\bf H$^m_0$} {\rm (i)} implies that for $t\in\R$
\begin{equation} \label{NGTEquation1}
\lim_{\epsi\to 0} \Big\|  U^\epsi (t) -  U^\epsi_{\rm diag} (t)
 \Big\|_{\mathcal{L}(\mathcal{H})} = 0\,,
\end{equation}
\item  {\bf  H$^m_0$} {\rm (ii)} implies that for some constant $C<\infty$ and all $t\in\R$
\begin{equation} \label{NGTEquation2}
 \Big\|  U^\epsi (t) -  U^\epsi_{\rm diag} (t)
 \Big\|_{\mathcal{L}(\mathcal{H})} \leq \,C\,\eta(\epsi^\frac{1}{2})\,(1+|t|)\,,
\end{equation}
\item  {\bf  H$^m_0$} {\rm (iii)} implies that for some constant $C<\infty$ and all $t\in\R$
\begin{equation} \label{NGTEquation3}
 \Big\|  U^\epsi (t) -  U^\epsi_{\rm diag} (t)
 \Big\|_{\mathcal{L}(\mathcal{H})} \leq \,C\,\eta(\epsi)\,(1+|t|)\,.
\end{equation}
\end{itemize}
\end{thm}

Note that in Theorem \ref{NoGapTheorem} the whole spectrum of possible rates of convergence
between $o(1)$ and $\mathcal{O}(\epsi)$ as in the case with gap is covered.
 The estimates for
the massless Nelson model as an application of Theorem \ref{NoGapTheorem} will
show that, in principle, all rates can occur.

The following proposition  shows that, assuming the first
part of {\bf H$^m_0$} but neither (i), (ii) or (iii), then
Assumption {\bf H$^m_0$} (i) always holds pointwise in $x$ if $P_*(x)$
is the spectral projection and has finite rank. The proof is analogous
to the one of Lemma 4 in \cite{AE}.

\begin{prop} \label{PropAE}
Assume {\bf H$^1_0$} without (i), (ii) or (iii). If $P_*(x)$ is the spectral projection
of $H_0(x)$ corresponding to the eigenvalue $E(x)$ and has finite rank, then
\begin{equation}\label{PE1}
\lim_{\delta\to 0} \|\,\delta\,R_{E(x)-i\delta}(H_0(x))\,
(\nabla_x P_*)(x)P_*(x)\|_{\mathcal{L(H}_{\rm f})}=0\,.
\end{equation}
\end{prop}

\begin{proof}
Since $P_*(x)$ has finite rank, the uniform statement (\ref{PE1}) follows
 if we can show that
$\lim_{\delta\to 0} \|\,\delta\,R_{E(x)-i\delta}(H_0(x))\,\psi \|=0$
for all $\psi\in \mbox{Ran}(\nabla_xP_*)(x)P_*(x)$. We have
\begin{equation} \label{limmu}
\lim_{\delta\to 0} \|\,  i\,\delta  R_{E(x)-i\delta}(H_0(x))\,\psi\|^2_{\mathcal{H}_{\rm f}}
= \lim_{\delta\to 0}\int_\R\,\mu_\psi(d\lambda)\,\frac{\delta^2}{(\lambda-E(x))^2 + \delta^2}= \mu_\psi(E(x))\,,
\end{equation}
where $\mu_\psi$ denotes the spectral measure of $H_0(x)$ for $\psi$.
Since  $P_*(x)$ is the spectral projection on  $\{E(x)\}$ and since,
 according to (\ref{gradPod}), Ran$(\nabla_xP_*)(x)P_*(x)\subset$ Ran$P_*^\perp(x)$ we have
 $\mu_\psi(E(x))=0$ and thus  (\ref{PE1}).
\end{proof}

It is clear from (\ref{limmu}) that additional information on the regularity of
the spectral measure $\mu_\psi$ provides some control on the rate of convergence
in  (\ref{limmu}). E.g., if  $\mu_\psi(d\lambda)=\rho_\psi(\lambda)d\lambda$
with $\rho_\psi\in L^\infty(\R,d\lambda)$, then
\[
\int_\R\,\mu_\psi(d\lambda)\,\frac{\delta^2}{(\lambda-E(x))^2 + \delta^2} \,\leq\,
\|\rho_\psi\|_\infty \,\int_\R\,d\lambda\,\frac{\delta^2}{(\lambda-E(x))^2 + \delta^2}
= \mathcal{O}(\delta)
\]
and hence (\ref{NGB1}) would hold pointwise in $x$ with
$\eta(\delta) = \delta^{1/2}$. In a sense, the rate $\mathcal{O}(\epsi\sqrt{\ln( 1/\epsi)})$
for the massless Nelson model (\ref{result}) is a consequence of the relevant spectral measure
having a density $\rho(\lambda) \sim \lambda-E(x)$.

We emphasize that  (\ref{PE1}) for all $x\in M$ does {\em not}
imply {\bf H$^m_0$} (i), even in the case of compact $M$. This is
because  for pointwise convergence to imply uniform convergence
one would need uniform equicontinuity of a sequence of functions.
However, in the time-adiabatic setting it is indeed sufficient to
have (\ref{PE1}) for almost all $x\in I$, where $I\subset \R$ is
the relevant time interval, see \cite{Note}.

\subsection{Proof of  Theorem \ref{NoGapTheorem}}

We start with the standard argument and find that on $D(H^\epsi)$
\begin{eqnarray} \label{Gap1}
 U^\epsi (t) -  U^\epsi_{\rm diag} (t)
& = &
-\, U^\epsi (t)\int_0^t\,
ds\,\frac{d}{ds} \left(  U^\epsi (-s)\,  U^\epsi_{\rm diag} (s)\right) \nonumber\\
&=&-\,\frac{i}{\epsi}\, U^\epsi (t)\int_0^t\,
ds\,  U^\epsi (-s)\,\big( H^\epsi \, - \,  H^\epsi_{\rm diag} \big)\,
 U^\epsi_{\rm diag} (s)\,,\nonumber
\end{eqnarray}
where
\begin{equation} \label{Gap3}
 H^\epsi \, - \,  H^\epsi_{\rm diag}  =
 P_*^\perp \,h^\epsi
 \,P_* \,+\, P_*\,h^\epsi
\,P_*^\perp=
 P_*^\perp \, \big[
h^\epsi ,P_* \big]\,P_*\,+\,\,\,\mathrm{adj.}\,.
\end{equation}
In (\ref{Gap3}) and in the following ``$\pm$ adj.'' means that the adjoint
operator of the first term in a sum is added resp.\ subtracted.
Inserting {\bf h$^m$} (i) into (\ref{Gap3}) and the result back into
  (\ref{Gap1}) one obtains
\begin{eqnarray} \label{Gap2}
\lefteqn{
\left\|  U^\epsi (t) -  U^\epsi_{\rm diag} (t) \right\|_{\mathcal{L(H)}}\,=}\\
\nonumber\\
&=&\,\left\| \, \int_0^t\, ds\,  U^\epsi (-s)\,\left(
P_*^\perp\,(\nabla_x P_*) \,P_*\cdot (D h)^\epsi\,P_* \,+\,
\mbox{adj.}\right) U^\epsi_{\rm diag} (s)
\right\|_{\mathcal{L(H)}}
 +\,\mathcal{O}(\epsi)|t|\,. \nonumber
\end{eqnarray}
In (\ref{Gap2}) we also used that
\begin{equation}\label{gradPod}
(\nabla_xP_*)(x) = P_*^\perp(x)(\nabla_xP_*)(x)P_*(x) +  \,\,{\rm
adj.} \,,
\end{equation}
which follows from $(\nabla_x P_*)(x)=(\nabla_x P_*^2)(x) =
(\nabla_xP_*)(x)P_*(x) +P_*(x)  (\nabla_xP_*)(x)$.

The nontrivial part in adiabatic theorems is to show that also the
remaining term on the right hand side of (\ref{Gap2}) vanishes as
$\epsi\to 0$. Assuming a gap condition, the basic idea is to
express the integrand, which is $\mathcal{O}(1)$, as the
time-derivative of a function that is $\mathcal{O}(\epsi)$ plus a
remainder that is $\mathcal{O}(\epsi)$ and integrate, cf.\
\cite{TS,ST}. The key ingredient in this case would be the
operator
\begin{equation}\label{FGapDef}
F(x) = R_{E(x)}(H_0(x))\,(\nabla_x P_*)(x)\,P_*(x)\,,
\end{equation}
which is, according to (\ref{gradPod}), well defined and bounded
if the eigenvalue $E(x)$ is separated from the rest of the
spectrum of $H_0(x)$ by a gap and if $P_*(x)$ is spectral.

The definition (\ref{FGapDef}) is made to give $[H_0,F]=
 P_*^\perp\,\big(\nabla_x P_*\big)\,P_*$. However, in absence of a gap
(\ref{FGapDef}) is not well defined as an operator on
$\mathcal{H}_{\rm f}$ and, following \cite{Note}, we shift the
resolvent into the complex plane and define
\[
F_\delta(x) =  R_{E(x)-i\delta}(H_0(x))\,P_*^\perp(x)\,(\nabla_xP_*)(x)\,P_*(x)\,.
\]
One now obtains
\begin{equation} \label{HFCom}
\big[\,H_0(x)\,,\,F_\delta(x)\,\big] =
P_*^\perp(x)\,\big(\nabla_x P_*\big)(x)\,P_*(x)  + Y_\delta(x)
\end{equation}
with
\begin{equation}\label{Ylim}
 Y_\delta(x) = -\, i\,\delta
R_{E(x)-i\delta}(H_0(x))\,(\nabla_xP_*)(x)\,P_*(x)\,.
\end{equation}
Assumptions {\bf H}$^m_0$ (i), (ii) and (iii) each imply that
$\lim_{\delta\to 0} \|Y_\delta\|_{\mathcal{L(H)}}=0$. To see this recall
 that for $A(\cdot)\in L^\infty(M,\mathcal{L(H}_{\rm f}))$
one has
\[
\|A\|_\mathcal{L(H)} = {\rm ess}\sup_{\hspace{-.5cm} x\in\R^n} \|A(x)\|_{\mathcal{L(H}_{\rm f})}\,.
\]
Note that for better readability we omit the Euclidean norm $|\ldots|$ in the notation and
understand that $\|A\|$ always includes also the Euclidean norm if $A$ is an operator
with $n$ components.
Thus with (\ref{Ylim}) we can make the remainder in (\ref{HFCom}) arbitrarily small
by choosing $\delta$ small enough. However, for the time being we
let $\delta>0$ but carefully keep track of the dependence of all
errors on $\delta$.

By assumption  $H_0(\cdot)\in C^{m}_{\rm b}(M,
\mathcal{L}(\mathcal{D},\mathcal{H}_{\rm f}))$ and $P_*(\cdot)\in
C^{m+1}_{\rm b}(M, \mathcal{L}(\mathcal{H}_{\rm f}))$, which
implies  $F_\delta(\cdot)\in C^m_{\rm
b}(M,\mathcal{L}(\mathcal{H}_{\rm f})^{\oplus n})$ and hence, according to
{\bf h$^m$} (i),
\begin{equation}\label{Gap5}
\left\| \, \big[ h^\epsi ,F_\delta\, \big]\,\right\|_{\mathcal{L(H)}}\, \leq\,
 C\,\sum_{j=1}^m \,\epsi^j \sup_{|\alpha|=j}\,\|\,\partial^\alpha_x F_\delta
\,\|_{\mathcal{L(H)}} =:\, f_1(\epsi,\delta)   \,.
\end{equation}
Combining (\ref{HFCom}) and (\ref{Gap5}) we obtain
\begin{equation}\label{Gap6}
\big[\,H^\epsi,F_\delta\,\big]\,= \,
 P_*^\perp\,\big(\nabla_x P_*\big)\,P_*\,+
\mathcal{O}(\|Y_\delta\|,f_1(\epsi,\delta))\,,
\end{equation}
where in (\ref{Gap6}) and in the following
$\mathcal{O}(a,b,c,\ldots)$ stands for a sum of operators whose
norm in $\mathcal{L(H)}$ is bounded by a constant times
$a+b+c+\ldots$. Defining
\[
B_\delta = F_\delta \cdot(D h)^\epsi\,P_* - \mbox{adj.}\,,
\]
one finds with {\bf h$^m$} (ii) and
$f_2(\delta) = \| F_\delta \|_{\mathcal{L}(\mathcal{H},D(H_0))}$
that
\begin{eqnarray}\label{Gap7}
\big[ H^\epsi, B_\delta\, \big] &=& \big[H^\epsi, F_\delta
\,\big]\cdot(D h)^\epsi\,P_* \, + F_\delta\cdot\big[ H^\epsi,
 (D h)^\epsi \big]\,P_*\,+F_\delta\cdot (D h)^\epsi\,\big[H^\epsi   ,\,P_*\big]
 +\,\mbox{adj.}\nonumber\\
& = & P_*^\perp\,\big(\nabla_x P_*\big)\,P_*\cdot (D h)^\epsi
\,+\,\mbox{adj.}\,+
\,\mathcal{O}(\epsi,\|Y_\delta\|,f_1(\epsi,\delta), \epsi
f_2(\delta))\,.
\end{eqnarray}
Now the integrand in (\ref{Gap2}) can be written as the time-derivative of
\[
A_\delta(s) =-\, i\,\epsi\,U^\epsi(-s)\,B_\delta\,U^\epsi(s)\,,
\]
plus a remainder:
\begin{eqnarray}
\frac{d}{ds}A_\delta(s) &=&
 U^\epsi(-s)\,[H^\epsi, B_\delta ]\,  U^\epsi(s)\, \nonumber\\
& =&   U^\epsi(-s)\,\left( P_*^\perp\,\big(\nabla_x
P_*\big)\,P_*\cdot(D h)^\epsi\,P_* + \mbox{adj.}\right)
U^\epsi(s)\nonumber\\&&+
\,\mathcal{O}(\epsi,\|Y_\delta\|,f_1(\epsi,\delta), \epsi
f_2(\delta))\,. \label{Gap8}
\end{eqnarray}
Inserting (\ref{Gap8}) into (\ref{Gap2}) enables us to do integration by parts,
\begin{eqnarray} \label{Gap9}
\lefteqn{\hspace{-1cm}
\left\|  U^\epsi (t) -  U^\epsi_{\rm diag} (t) \right\|_\mathcal{L(H)}
\,\leq} \nonumber\\
&\leq&\left\|\, \int_0^t\, ds\, \left(
\frac{d}{ds}A_\delta(s)\right) U^\epsi (-s)\, U^\epsi_{\rm diag}
(s)\,\right\|_\mathcal{L(H)}\nonumber\\&& \hspace{2cm}+\,|t|\,
\mathcal{O}\Big(\epsi,\|Y_\delta\|,f_1(\epsi,\delta),
\epsi f_2(\delta)\Big)      \nonumber\\
&\leq& \|A_\delta(t)\|_\mathcal{L(H)}+\|A_\delta(0)\|_\mathcal{L(H)} \nonumber\\
&&+\left\|\,\int_0^t\, ds\, A_\delta (s)\left(\frac{d}{ds} U^\epsi
(-s)\, U^\epsi_{\rm diag}
(s)\right)\right\|_\mathcal{L(H)}\nonumber\\ &&+
\,|t|\,\mathcal{O} \Big(\epsi,\|Y_\delta\|,f_1(\epsi,\delta),
\epsi f_2(\delta)\Big)
\nonumber\\
&\leq&  C\,\epsi\,(2+|t|)\,\|F_\delta\|_{\mathcal{L(H)}} +
\,|t|\,\mathcal{O} \Big(\epsi,\|Y_\delta\|,f_1(\epsi,\delta),
\epsi f_2(\delta)\Big)  \,.\label{cbound}
\end{eqnarray}
For the last inequality in (\ref{cbound}) we used that
$\|A_\delta(t)\|_{\mathcal{L(H)}}\leq
C\,\epsi\|F_\delta\|_{\mathcal{L(H)}}$ uniformly for $t\in \R$
and that
\[
\frac{d}{ds}\,U^\epsi (t_0,s)\, U^\epsi_{\rm diag} (s,t_0) =
-\frac{i}{\epsi}\, U^\epsi (t_0,s)\,\big( H^\epsi(s) \, - \,  H^\epsi_{\rm diag}(s) \big)\,
 U^\epsi_{\rm diag} (s,t_0)
\]
is bounded uniformly, according to (\ref{Gap3}) and {\bf h$^m$} (i).

Writing out the various terms in (\ref{cbound}) explicitly, we conclude
 that there is a constant $C<\infty$ such that
\begin{eqnarray} \label{BOUND}
\lefteqn{\hspace{-1cm} \left\|  U^\epsi (t) -  U^\epsi_{\rm diag}
(t) \right\|_\mathcal{L(H)} \,\leq
C\,\epsi\,\|F_\delta\|_\mathcal{L(H)}+ C\,|t|\,\Big(\epsi+
\|Y_\delta\|_\mathcal{L(H)} } \nonumber\\
&&  + \epsi\,\|F_\delta\|_{\mathcal{L(H},D(H_0))} +
 \epsi\,\|F_\delta\|_\mathcal{L(H)}
+ \sum_{j=1}^m\,\epsi^j \sup_{|\alpha|=j} \|\partial^\alpha_x F_\delta
\|_\mathcal{L(H)}
\Big)\,.
\end{eqnarray}
Hence we are left to establish
bounds on  $F_\delta $, on its derivatives and on $Y_\delta$ in
 terms of $\delta$, which is the content of the following Lemma.

\begin{lem}\label{BDLem}  Assume {\bf H$^m_0$}, then
$F_\delta(\cdot)\in C^m_{\rm b}(M,\mathcal{L}(\mathcal{H}_{\rm f})^{\oplus n})$
and there is a constant $C<\infty$ such that for $\delta\in(0,\delta_0]$
\begin{equation} \label{BDbounds}
 \|F_\delta\|_{\mathcal{L}(\mathcal{H},D(H_0))}\leq\,  \frac{C}{\delta}\,\eta(\delta)
\,,
\end{equation}
\begin{equation} \label{BDbounds2}
 \sup_{|\alpha|=j}\,\|\,\partial^\alpha_x\,F_\delta\,\|_{\mathcal{L}(\mathcal{H})}
\leq \,\frac{C}{\delta^{j+1}}\,\eta(\delta)\quad \mbox{for}\,\,1\leq j\leq m \,.
\end{equation}
In case  {\bf H$^m_0$} {\rm (i)} holds, we have (\ref{BDbounds}) and  (\ref{BDbounds2})
with $\eta(\delta)=1$.
Furthermore,
if {\bf H$^m_0$} {\rm (i)} holds, then
$\lim_{\delta\to 0} \|Y_\delta\|_\mathcal{L(H)} = 0$ and if  {\bf H$^m_0$}
{\rm (ii)} or {\rm (iii)} holds, then
$ \|Y_\delta\|_\mathcal{L(H)} \,\leq \,C\,\eta(\delta)$.

If  {\bf H$^m_0$} (iii) holds, then (\ref{BDbounds2}) can be improved to
\begin{equation} \label{BDbounds3}
 \sup_{|\alpha|=j}\,\|\,\partial^\alpha_x\,F_\delta\,\|_{\mathcal{L}(\mathcal{H})}
\leq \,\frac{C}{\delta^{j}}\,\eta(\delta)\quad \mbox{for}\,\,1\leq j\leq m \,.
\end{equation}
\end{lem}

Before we turn to the proof of Lemma \ref{BDLem} we finish the proof of Theorem \ref{NoGapTheorem}.
Assuming {\bf H$^m_0$} (i),
(\ref{NGTEquation1}) follows by inserting the bounds from Lemma
 \ref{BDLem} into (\ref{BOUND}) and choosing $\delta = \delta(\epsi)$ such that $\lim_{\epsi\to 0}
\delta(\epsi) = 0$ and  $\lim_{\epsi\to 0} \epsi/\delta(\epsi)^2 = 0$.

If  {\bf H$^m_0$} (ii) holds, then the bounds (\ref{BDbounds}) and
(\ref{BDbounds2}) inserted into (\ref{BOUND}) yield
\begin{equation}\label{BOUND2}
\left\|  U^\epsi (t) -  U^\epsi_{\rm diag} (t)
\right\|_\mathcal{L(H)} \,\leq C\,
\epsi\,\frac{\eta(\delta)}{\delta} +C\,\Big( \epsi+ \eta(\delta)
+ \epsi\,\frac{\eta(\delta)}{\delta} + \sum_{j=1}^m \,\epsi^j\,
\frac{\eta(\delta)}{\delta^{j+1}} \Big)|t|\,.
\end{equation}
In (\ref{BOUND2}) the optimal choice is $\delta(\epsi) = \epsi^\frac{1}{2}$,
which gives (\ref{NGTEquation2}).
Finally, the bounds (\ref{BDbounds}) and  (\ref{BDbounds3}) inserted into (\ref{BOUND})
yield
\[
\left\|  U^\epsi (t) -  U^\epsi_{\rm diag} (t)
\right\|_\mathcal{L(H)} \,\leq C\,
\epsi\,\frac{\eta(\delta)}{\delta} +C\,\Big(\epsi+  \eta(\delta)
+ \epsi\,\frac{\eta(\delta)}{\delta} + \sum_{j=1}^m \,\epsi^j\,
\frac{\eta(\delta)}{\delta^{j}} \Big)|t|\,,
\]
where  the optimal choice $\delta(\epsi) = \epsi$ gives (\ref{NGTEquation3}).

\begin{proof}[Proof of Lemma \ref{BDLem}]
We abbreviate  $R_{E(x)-i\delta}(H_0(x))$ as $R(\delta,x)$ in this proof and note that
 $R(\delta,\cdot)\in C^m_{\rm b}(M,\mathcal{L}(\mathcal{H}_{\rm f}))$
and thus $F_\delta(\cdot) \in C^m_{\rm b}(M,\mathcal{L}(\mathcal{H}_{\rm f})^{\oplus n})$
follow from
$H_0(\cdot)\in C^m_{\rm b}(M, \mathcal{L}(\mathcal{D},\mathcal{H}_{\rm f}))$
together with  $P_*(\cdot)\in C^{m+1}_{\rm b}(M,\mathcal{L}(\mathcal{H}_{\rm f}))$.

We start with the case {\bf H$^m_0$} (ii), where (i) is included by making the
obvious changes for $\eta(\delta)=1$.
Assumption  {\bf H$^m_0$} (ii) immediately yields
\begin{equation}\label{fbound}
 \|F_\delta\|_{\mathcal{L}(\mathcal{H})}
\leq\, \frac{C}{\delta}\,\eta(\delta)
\end{equation}
and  the bound on $Y_\delta$.
 (\ref{BDbounds}) follows from
$H_0(x)R(\delta,x) = {\bf 1} + (E(x) -i\delta)R(\delta,x)$ and (\ref{fbound})
together with the assumption that $E(x)$ is uniformly bounded.

 For (\ref{BDbounds2}) we start by  observing that
\begin{eqnarray} \label{gradRD}
\nabla_x F_\delta(x)& =& R(\delta,x)\,\nabla_x \Big( (\nabla_x
P_*)(x)P_*(x) \Big) +
\Big(\nabla_x R(\delta,x) \Big) (\nabla_x P_*)(x)P_*(x) \\
& = & R(\delta,x)\nabla_x \Big( (\nabla_x P_*)(x)P_*(x) \Big)
 - R(\delta,x)(\nabla_x H_0(x) - \nabla_x E(x)) F_\delta(x). \nonumber
\end{eqnarray}
Using (\ref{fbound}) and the fact that $|\nabla_x E(x)|$ is uniformly bounded by
assumption, we infer from (\ref{gradRD}) that
\[
\|\,|\nabla_x F_\delta|\,\|_\mathcal{L(H)} =
\sup_{x\in M} \|\,|\nabla_x F_\delta(x)|\, \|_{\mathcal{L(H}_{\rm f})}\leq
C(\delta^{-1} + \delta^{-2}\eta(\delta))\,.
\]
 Hence  (\ref{BDbounds2}) follows for $j=1$,
since $\delta\leq \eta(\delta)$ by assumption.

By differentiating (\ref{gradRD}) again, we find, using a reduced notation with
obvious meaning, that
\begin{equation} \label{Nab2F}
\nabla^{(2)} F_\delta =  - 2 R_\delta(\nabla H_0 - \nabla E)\,\nabla F_\delta +
 R_\delta\,\nabla^{(2)}\Big( (\nabla P_*)P_* \Big)
- R_\delta(\nabla^{(2)} H_0 - \nabla^{(2)} E)\, F_\delta\,.
\end{equation}
Hence  $\|\nabla^{(2)} F_\delta \|_{\mathcal{L(H)}}\leq
C(\delta^{-3}\eta(\delta)+\delta^{-1} + \delta^{-2}\eta(\delta))$ which proves
 (\ref{BDbounds2}) for $j=2$. By repeated differentiation one finds inductively
(\ref{BDbounds2}) for $j\leq m$.

To show (\ref{BDbounds3}) assuming {\bf H$^m_0$} (iii), note that
 (\ref{BDbounds3}) holds by assumption for $j=1$ and
inserted into (\ref{Nab2F}) it gives
$\|\nabla^{(2)} F_\delta \|_{\mathcal{L(H)}}\leq
C(\delta^{-2}\eta(\delta)+\delta^{-1} + \delta^{-2}\eta(\delta))$.
Analogously the estimates for all larger $j\leq m$ are improved by
a factor of $\delta$.
\end{proof}

\section{Effective dynamics for the massless Nelson model} \label{SAN}

As explained in the introduction,
we consider $N$ spinless particles coupled to a scalar, massless,  Bose field
with an ultraviolet regularization in the interaction.
This class of models is nowadays called Nelson's model \cite{Arai,Volker,LMS2,LMS}
after E.\ Nelson \cite{Nelson}, who studied the ultraviolet problem.
We briefly complete the introduction of the model and collect some basic, well known facts.

A point in the configuration space  $\R^{3N}$ of the particles is denoted by
$x=(x_1,\ldots,x_N)$ and the Hamiltonian $H^\epsi_{\rm p}$ for the particles
is defined in (\ref{Hpepsi}).
$H^\epsi_{\rm p}$ is self-adjoint on the domain $H^1(\R^{3N})$, the first Sobolev space.

The Hilbert space for the scalar field is the bosonic Fock space over
$L^2(\R^3)$ defined in (\ref{Fock}).
On $D(\sqrt{\mathcal{N}})$, $\mathcal{N}$ the number operator,
the annihilation operator $a(f)$ acts for $f\in L^2(\R^3)$ as
\[
(a(f)\psi)^{(m)}(k_1,\ldots,k_m) = \sqrt{m+1}\int_{\R^3}\,dk\,
\bar f(k)\,\psi^{(m+1)}(k,k_1,\ldots,k_m)\,,
\]
where $\psi=(\psi^{(0)},\psi^{(1)},\psi^{(2)},\ldots)\in D(\sqrt{\mathcal{N}})$
if and only if $\sum_{m=0}^\infty m\|\psi^{(m)}\|^2<\infty$.
The adjoint $a^*(f)$, which  is also defined on $D(\sqrt{\mathcal{N}})$, is the
creation operator and for $f,g\in L^2(\R^3)$ the operators
$a(f)$ and $a^*(g)$ obey the canonical commutation relations (CCRs)
\begin{equation} \label{CCR}
[ a(f),a^*(g) ] = \int_{\R^3}\,dk\,\bar f(k)\,g(k)=:
\langle f,g\rangle\,,\quad[ a(f),a(g) ]=[ a^*(f),a^*(g) ]=0\,.
\end{equation}
It is common to write $a(f)= \int_{\R^3}\,dk\,\bar f(k)\,a(k)$.
The Hamiltonian of the field as defined in (\ref{Hf}) can  formally be written as
\[
H_{\rm f} = \int_{\R^3} \,dk\,|k|\,a^*(k)\,a(k)\,.
\]
More explicitly, on the $m$-particle sector the action of  $H_{\rm f}$
is
 $(H_{\rm f}\psi)^{(m)}(k_1,\ldots,k_m) =
\sum_{j=1}^m |k_j| \psi^{(m)}(k_1,\ldots,k_m)$, and $H_{\rm f}$
is self-adjoint on its maximal domain.
 For $f\in L^2(\R^3)$  the Segal field operator
\[
\Phi(f) = \frac{1}{\sqrt{2}}\Big( a(f) + a^*(f) \Big)
\]
 is essentially self-adjoint on  $D(\sqrt{\mathcal{N}})$. The field operator
$\phi$ as used in (\ref{fieldop}) is related to $\Phi$ through
$\phi(f) = \Phi(f/\sqrt{|k|})$.
For the following it turns out to be more convenient to write the
interaction  Hamiltonian in terms of $\Phi$, where
\[
H_{\rm I} =  \Phi\Big( |k|\,v(x,k) \Big)
\]
 acts on the Hilbert space  $\mathcal{H}= L^2(\R^{3N})\otimes\mathcal{F}$
of the full system.
We will consider two different choices for $v(x,k)$ in more detail. For the
standard Nelson model (SN), as discussed in the introduction,
one has
\begin{equation}\label{vsnDef}
v_{\rm SN}(x,k) =\sum_{j=1}^N e^{ik\cdot x_j} \frac{\hat \rho_j(k)}{|k|^{3/2}}\,.
\end{equation}
For the infrared-renormalized models (IR), as considered
by Arai \cite{Arai} and, more generally, by L\"orinczi, Minlos and Spohn
\cite{LMS}, one has
\begin{equation}\label{virDef}
v_{\rm IR}(x,k) =\sum_{j=1}^N \left(e^{ik\cdot x_j}-1\right)
\frac{\hat \rho_j(k)}{|k|^{3/2}}\,.
\end{equation}
In both cases, the  charge distribution $\rho_j\in L^1(\R^3)$ of the $j^{th}$ particle
is assumed to be real-valued and spherically symmetric. As to be discussed below,
cf.\ Remarks \ref{IRRem} and \ref{CutRem},
we have to  assume the infrared condition (\ref{IRCon})
for the (SN) model, but {\em not} for the (IR) model. The infrared condition
implies, in particular, that the total charge of the system of $N$ particles must be zero.

The full Hamiltonian is given as the sum
\begin{equation} \label{HNSDEF}
H^\epsi= H^\epsi_{\rm p}\otimes {\bf 1} + {\bf 1}\otimes H_{\rm f}
+ H_{\rm I} + V_{\rm IR}\otimes {\bf 1}
\end{equation}
and is essentially self-adjoint on $D(H_{\rm p}\otimes{\bf 1})\cap
D({\bf 1}\otimes H_{\rm f})$
if $\sup_x \|\,|k|^s v(x,k)\,\|_{L^2(\R^3)}<\infty $ for $s\in\{\frac{1}{2},1\}$.
{\em Only} in the (IR) model a potential
$V_{\rm IR}$ is added, which acts as multiplication with the bounded,
real-valued function
\begin{equation}\label{VIRDef}
V_{\rm IR}(x) = \sum_{j,i=1}^N\int_{\R^3} dk\,
\frac{\hat\rho_j(k)\hat\rho_i(k)^*}{|k|^2}e^{-ik\cdot x_j} +\frac{1}{2}
\int_{\R^3} dk\,\left| \frac{\sum_{j=1}^N\hat\rho_j(k)}{|k|}\right|^2\,.
\end{equation}

\begin{rem}\label{IRRem}
If the charge distributions satisfy the infrared condition (\ref{IRCon}), then
the (SN) Hamiltonian and the (IR) Hamiltonian are related by the unitary
transformation
\[
U_{\rm G} = \exp\left( -i\Phi\left( i\sum_{j=1}^N\frac{\hat\rho_j^{\,*}(k)}{|k|^{3/2}}
\right)\right)\,,
\]
cf.\ \cite{Arai}, which is related to the Gross transformation \cite{Nelson} for $x=0$.
If  the infrared condition is not satisfied, the (SN) model and the (IR) model
carry two inequivalent representations of the CCRs for the field operators.
Physically speaking, the transformation $U_{\rm G}$ removes the mean field
that the $N$ charges would generate, if all of them would be moved to the origin.
The vacuum in the (IR) representation corresponds to this removed mean field in
the original representation, a fact which has to be taken care of in the interaction:
 for each particle the interaction term is now evaluated relative to the
interaction at $x=0$, cf.\ (\ref{virDef}), which makes also
necessary the counter terms $V_{\rm IR}$. If the total charge of
the system is different from zero, then the mean field is long
range and, as a consequence, the corresponding transformation is
no longer unitarily implementable. Indeed, it was shown
that the (SN) Hamiltonian with confining potential does {\em not} have a ground
state, cf.\ \cite{LMS2}, while the (IR) Hamiltonian with the same confining
potential does have a ground state, cf.\ \cite{Arai}.
\er\end{rem}

In order to apply Theorem \ref{NoGapTheorem} we observe that $H_{\rm I}(x)$ acts
for fixed $x\in\R^{3N}\,(\cong M)$ on
$\mathcal{F}\,(\cong\mathcal{H}_{\rm f}) $ and with  $H_0(x) = H_{\rm f}
+ H_{\rm I}(x)+ V_{\rm IR}(x)$ we have
\[
H^\epsi= H^\epsi_{\rm p}\otimes {\bf 1} + \int_M^\oplus \,dx\,H_0(x)
\quad\Big(\, \cong\, h^\epsi + H_0\,\Big)\,.
\]

The following proposition collects some results about $H_0(x)$ and
its ground state. Its proof
is postponed to after the presentation of the the main theorem.

\begin{prop}\label{NM1Lem}
Assume that $v(x,\cdot)\in L^2(\R^{3})$ for all $x\in\R^{3N}$ and that for some $n\geq 1$
\begin{enumerate}
\item $|\cdot|\partial^\alpha_x v(x,\cdot)\in L^2(\R^{3})$ for all $x\in\R^{3N}$
and $0\leq|\alpha|\leq n$,
\item $\sup_{x\in\R^{3N}} \|\,\sqrt{|\cdot|}\,
\partial^\alpha_x v(x,\cdot)\,\|_{L^2(\R^3)} <\infty$ for  $0\leq|\alpha|\leq n$,
\item $\sup_{x\in\R^{3N}} \|\,
\partial^\alpha_x v(x,\cdot)\,\|_{L^2(\R^3)} <\infty$ for  $1\leq|\alpha|\leq n$.
\end{enumerate}
Let {\rm Im}$\langle v(x,\cdot),\nabla_x v(x,\cdot)\rangle_{L^2(\R^3)}=0$ for
all $x\in\R^{3N}$ and $V_{\rm IR}(\cdot)\in C^3_{\rm b}(\R^{3N})$. Then
\begin{itemize}
\item[1.]
$H_0(x)$ is self adjoint on $\mathcal{D} =D(H_{\rm f})$ for all $x\in  \R^{3N}$ and
$H_0(\cdot)\in C^n_{\rm b}( \R^{3N},$ $\mathcal{L}(\mathcal{D},\mathcal{F}))$, where
$\mathcal{D}$ is equipped with the graph-norm of $H_{\rm f}$.
\item[2.]
$H_0(x)$ has a unique ground state $\Omega(x)$ for all $x\in \R^{3N}$ and, in particular,\\
 $H_0(x)\Omega(x)=E(x)\Omega(x)$ for
\begin{equation}\label{EGSDef}
E(x) =   - \frac{1}{2}\int_{\R^3}\,dk\,|k|\,|v(x,k)|^2 + V_{\rm IR}(x).
\end{equation}
Furthermore $\Omega(\cdot)\in C^n_{\rm b}( \R^{3N}, \mathcal{F})$.
\end{itemize}
\end{prop}

It is straightforward to check that Im$\langle v(x,\cdot),\nabla_x
v(x,\cdot)\rangle_{L^2(\R^3)}=0$
for $v_{\rm SN}$ defined in (\ref{vsnDef}) and $v_{\rm IR}$
defined in (\ref{virDef}).
For the (SN) model as well as for the (IR) model  (\ref{EGSDef}) is easily evaluated
and one finds $E(x)=E_0(x)$ as given in (\ref{E0Def}).

\begin{rem}\label{CutRem}
For the (SN) model the assumptions made on $v(x,k)$ in Proposition
\ref{NM1Lem} are satisfied, if  $\hat \rho_j(k)$
decays sufficiently fast for large $|k|$ and each $j=1,\ldots, N$,
or, equivalently, if $\rho_j(x)$ is sufficiently smooth. This is an
ultraviolet condition individually for each particle. But
$v(x,\cdot)\in L^2(\R^{3})$ follows from $v(0,\cdot)\in L^2(\R^{3})$,
which is exactly the global infrared
condition (\ref{IRCon}).

While the necessity for an ultraviolet
regularization remains in the (IR) model, the infrared condition
is replaced by $\sum_j |k|^{-\frac{3}{2}}
\widehat\rho_j(k)(e^{ik\cdot x_j}-1) \in L^2(\R^3)$, which can be satisfied
without having $\sum_j \widehat\rho_j(0) = 0$. Thus the (IR) model allows us
to consider particles with total charge different from zero.
\er\end{rem}

Let
$P_*(x) = |\Omega(x)\rangle \langle\Omega(x)| $, then $P_*(\cdot) \in  C^n_{\rm b}(\R^{3N},
\mathcal{L}(\mathcal{F}))$
and Ran$P_*$ is a candidate for an adiabatically decoupled subspace.
Indeed, we will show that $h^\epsi = H^\epsi_{\rm p}\otimes{\bf 1}$ satisfies Assumption {\bf h$^m$}
with
$(Dh)^\epsi_{x_j} = -i\epsi\nabla_{x_j}/ \sqrt{ - \epsi^2 \Delta_{x_j} + 1}$ and that
$H_0$ and $P_*$ satisfy Assumption {\bf H$^m_0$} (iii) with
 $\eta(\delta)=\delta\sqrt{\ln(1/\delta)}$ if particles with charges different from zero are present
and $\eta(\delta)=\delta$ if all particles have total charge zero.
Hence we can apply Theorem \ref{NoGapTheorem} to conclude that for some constant $C<\infty$
\begin{equation}\label{HHdiagN}
\left\|  e^{-iH^\epsi t/\epsi}  -  e^{-iH^\epsi_{\rm diag} t/\epsi} \right\| \leq \, C\,\eta(\epsi)
\,(1+|t|)\,,
\end{equation}
with $H^\epsi_{\rm diag} = P_*\,H^\epsi\,P_* + ({\bf 1}-P_*)\,H^\epsi\, ({\bf 1}-P_*)$.

Next observe that the ground state band subspace Ran$P_*$ is unitarily
equivalent to the Hilbert space $L^2( \R^{3N})$ of the $N$ particles in a natural way. Let
\begin{equation}\label{UCDef}
\mathcal{U}: {\rm Ran}P_* \to L^2(\R^{3N})\,,\quad \psi \mapsto  (\mathcal{U}\psi)(x) =
 \langle \Omega(x),\psi(x) \rangle_\mathcal{F}\,,
\end{equation}
then it is easily checked that for $\varphi\in L^2( \R^{3N})$
\[
\mathcal{U}^*\varphi = \mathcal{U}^{-1}\varphi = \int_{\R^{3N}}^\oplus\,dx\,\varphi(x)\,\Omega(x)\,.
\]
Hence the part of $H^\epsi_{\rm diag}$ acting on Ran$P_*$ is unitarily equivalent to
\[
\widetilde H^\epsi_{\rm eff}= \mathcal{U} \,P_*\,H^\epsi\,P_*\,\mathcal{U}^*
\]
 acting on the
Hilbert space of the $N$ particles {\em only}.
The following theorem shows that $\widetilde H^\epsi_{\rm eff}$ has, at leading order,
exactly the form expected from the heuristic ``Peierls substitution'' argument.

\begin{thm} \label{NelsonTheorem} Let $H^\epsi$ be defined as in (\ref{HNSDEF}) with
 $v(x,k)$  either  $v_{\rm SN}(x,k)$ as in  (\ref{vsnDef}) or
$v_{\rm IR}(x,k)$ as in (\ref{virDef}).
For $1\leq j\leq N$ let $\rho_j\in L^1(\R^3)$ such that
$\hat \rho_j$ satisfies $|k|^s\widehat\rho_j(k)\in L^2(\R^3)$ for
$s\in \{-1,4\}$. For the (SN) model assume, in addition,
the infrared condition (\ref{IRCon}). Let
\[
H_{\rm eff}^\epsi = H^\epsi_{\rm p} + \sum_{j=2}^N\sum_{i=1}^{j-1} V_{ij}(x_i - x_j) + e_0\,,
\]
with $V_{ij}(z)$ and $e_0$ as in (\ref{vijDef}) and (\ref{e0Def}). Then there is a constant
$C<\infty$ such that
\begin{equation} \label{NTEquation}
\Big\| \Big(  e^{-iH^\epsi t/\epsi}  -\mathcal{U}^*\,  e^{-iH^\epsi_{\rm eff} t/\epsi}
\,\mathcal{U} \Big) P_* \Big\| \leq \, C\,\eta(\epsi)\,(1+|t|)\,,
\end{equation}
where  $\eta(\epsi)=\epsi\sqrt{\ln(1/\epsi)}$. If all charges satisfy the infrared condition
individually, i.e.\ if $\hat\rho_j(k)/|k|^{3/2}\in L^2(\R^3)$ for all $j=1,\ldots,N$,
then (\ref{NTEquation}) holds with $\eta(\epsi)=\epsi$.
\end{thm}

For sake of better readability the global energy shift $e_0$ was,
as opposed to (\ref{result}), absorbed into the definition of $H_{\rm eff}^\epsi$.

\begin{rem}
The unitary $\mathcal{U}$ intertwines the position operator $x\otimes{\bf 1}$ on
$L^2(\R^{3N})\otimes\mathcal{F}$ with the position operator $x$ on $L^2(\R^{3N})$ exactly
and the momentum operator $-i\epsi\nabla_x\otimes{\bf 1}$ on $L^2(\R^{3N})\otimes\mathcal{F}$
with $-i\epsi\nabla_x$ on $L^2(\R^{3N})$ up to an error of order $\epsi$.
Thus one can directly read off the position distribution and approximately also the
momentum distribution  of the particles
from the solution $\psi(t)= e^{-iH^\epsi_{\rm eff} t/\epsi}\psi_0$ of the effective dynamics.
It is {\em not} necessary to transform back to the full Hilbert space using $\mathcal{U}^*$.
\er\end{rem}

\begin{rem}\label{leakrem}
From the discussion of the introduction
one expects, on physical grounds, that the energy lost through
radiation is of order $\mathcal{O}(\epsi^3)$
after times of order $\mathcal{O}(\epsi^{-1})$. Hence the error of order 
$\mathcal{O}(\epsi\sqrt{\ln 1/\epsi})$
in (\ref{NTEquation}) is not optimal in the sense that the error does not correspond
to emission of free bosons and thus to dissipation.
Indeed, we expect that the situation is similar
to adiabatic perturbation theory with gap, cf.\ \cite{PST}. There should be a subspace
Ran$P^\epsi_*$ which is $\epsi$-close to Ran$P_*$ and for which the analogous expression to
(\ref{HHdiagN}) holds with an error of order $\mathcal{O}(\epsi^\frac{3}{2})$, possibly with
a logarithmic correction.
The corresponding effective Hamiltonian would then contain two additional terms of order
$\epsi^2$, which, for the case of quadratic dispersion
\[
H^\epsi_{\rm p} = -\,\sum_{j=1}^N \frac{\epsi^2}{2}\Delta_{x_j}
\]
for the particles, can be calculated
using formula (47) in \cite{PST}. As result we obtain
for the Weyl symbol of the effective Hamiltonian including the momentum dependent
Darwin term
\begin{equation}\label{Darwin}
H_{\rm eff}(p,q) = \sum_{j=1}^N \frac{1}{2m_j^\epsi}\, p_j^2 + E(q)
+ \frac{\epsi^2}{2}\, \sum_{j<i} \int_{\R^3} dk\,\frac{(p_j\cdot \kappa)(p_i\cdot\kappa)}{|k|^2}
e^{-ik\cdot(q_j-q_i)} \hat\rho_j^*(k)\hat\rho_i(k)
\end{equation}
with $m_j^\epsi =1/(1 + \frac{\epsi^2}{2} e_j)$ and
\[
e_j = \frac{1}{4\pi} \int_{\R^3\times\R^3}dv\,dw\,\frac{\rho_j(v)\rho_j(w)}{|v-w|}
\]
the electromagnetic mass.
As explained above, for the rigorous justification of (\ref{Darwin})
a space-adiabatic theorem without gap but for rotated subspaces $P_*^\epsi$
is needed and thus it is beyond the scope of the present paper.
\er\end{rem}

Before proving Theorem \ref{NelsonTheorem} we make up for the

\begin{proof}[Proof of Proposition \ref{NM1Lem}] A
standard estimate (cf.\ e.g.\ \cite{Volker} Proposition 1.3.8)
shows that for $f\in L^2(\R^3)$ and any $a>0$
\begin{equation} \label{phiest}
\|\Phi(f)\psi\|^2_\mathcal{F} \leq a \|H_{\rm
f}\psi\|^2_\mathcal{F} +
\left(\frac{\|f/\sqrt{|\cdot|}\,\|^4_{L^2(\R^3)}}{a} +
2\right)\|\psi\|^2_\mathcal{F}\,.
\end{equation}
Hence  $\Phi(f)$ is infinitesimally
$H_{\rm f}$-bounded whenever $\|f\|_{L^2(\R^3)}+
\|f/\sqrt{|\cdot|}\|_{L^2(\R^3)}<\infty$. Then Kato-Rellich implies
that $H_0(x)$ is self-adjoint on $D(H_{\rm f})$, since
$\sqrt{|\cdot|}v(x,\cdot)\in L^2$ is assumed.
Using (i), (ii)
and $V_{\rm IR}(\cdot)\in C^n_{\rm b}(\R^{3N})$, we obtain from  (\ref{phiest}) that
\begin{equation}\label{NabH}
\partial^\alpha_x  H_0(x) = \Phi\left( |k|\,\partial^\alpha_x v(x,k)  \right) + \partial^\alpha_x V_{\rm IR}(x)
\end{equation}
is relatively bounded with respect to $H_{\rm f}$ for $|\alpha|\leq n$.
Moreover, (ii), (\ref{phiest}) and (\ref{NabH})  imply that
$H_0(\cdot)\in C^n_{\rm b}( \R^{3N},\mathcal{L}(\mathcal{D},\mathcal{F}))$.

To compute the ground state energy $E(x)$ observe that from
``completing the square'' one finds
\[
H_0(x) = \int_{\R^3}dk\,|k|\, \Big(a^*(k) +\frac{ v^*(k,x)}{\sqrt{2}}\Big)
\Big(a(k) + \frac{ v(x,k)}{\sqrt{2}} \Big)  - \frac{1}{2}\int_{\R^3}dk\,|k|\,|v(x,k)|^2 +
V_{\rm IR}(x)\,.
\]
It is well known that the map $a(k) \mapsto a(k) + v(x,k)/\sqrt{2}$ comes from the
 unitary transformation $U(x) = \exp\Big(i\Phi\big(iv(x,\cdot)\big)\Big)$, i.e.
\begin{equation}\label{Bog}
U(x)\,a(f)\,U^*(x) = a(f) + \langle f,v(x,\cdot)\rangle\,,
\end{equation}
whenever $v(x,\cdot)\in L^2$. Equation (\ref{Bog}) follows from  the fact that
 $[A,[A,B]]=0$ implies
 $[\exp(iA),B] = \exp(iA)[iA,B]$ and the CCRs.
 Transformations of the form (\ref{Bog}) are  called Bogoliubov transformations.

 Therefore $H_0(x) = U(x)\,H_{\rm f}\,U^*(x)\, + E(x)$
with $ E(x) =  -\frac{1}{2} \int_{\R^3}\,dk\,|k|\,|v(x,k)|^2 +V_{\rm IR}(x)$.
Since $H_{\rm f}\, \Omega_0 = 0$ for
the unique ground state $\Omega_0 = (1,0,0,\ldots)\in \mathcal{F}$, we find
\[
H_0(x) \Omega(x) = E(x)\Omega(x) \quad\mbox{with}\quad \Omega(x) = U(x) \Omega_0\,.
\]

Next we need to take derivatives of $U(x)$ with respect to $x$.
It follows from the CCRs (\ref{CCR}) that for $f,g\in L^2(\R^3)$
\[
\big[ \Phi(f),\Phi(g) \big] = i \,{\rm Im}\langle f,g\rangle\,,
\]
and thus, by assumption, that
 $\nabla_x \Phi(iv(\cdot,x)) = \Phi(i\nabla_x v(\cdot,x))$ commutes with $ \Phi(iv(\cdot,x))$.
Hence we obtain that on $D(\sqrt{\mathcal{N}})$
\begin{equation} \label{nabu}
\nabla_x U(x) = U(x)\, i\Phi(i\nabla_x v(\cdot,x)) =
i\Phi(i\nabla_x v(\cdot,x))\, U(x)\,.
\end{equation}
 By further differentiating (\ref{nabu}) we can get up
to $n^{th}$ derivatives since $\partial_x^\alpha v(x,k)\in L^2(\R^3)$ for
$|\alpha|\leq n$. In particular
we find with (iii) that $\Omega(x)= U(x)\Omega_0 \in C^n_{\rm b}( \R^{3N},\mathcal{F})$.
\end{proof}

\begin{proof}[Proof of Theorem \ref{NelsonTheorem}]
We start by showing that the assumptions of Theorem
\ref{NoGapTheorem} are indeed satisfied and thus (\ref{HHdiagN})
follows.

It is straightforward to check that the assumptions on $\hat \rho_j$
imply the assumptions of  Proposition \ref{NM1Lem} for $n=5$.
Hence the first part of {\bf H}$_0^4$ follows with $P_*(x)= |\Omega(x)\rangle\langle
\Omega(x)|$.
For {\bf H}$_0^4$ (iii) observe that, using (\ref{nabu}),
$\nabla_x\Omega(x)= \nabla_x U(x)\Omega_0 = U(x)\,i\Phi(i\nabla_x v(\cdot,x))\Omega_0 $,
and thus
\begin{equation} \label{Omeg}
\langle\Omega(x),  \nabla_x\Omega(x)\rangle_\mathcal{F} =
\langle\Omega_0, i\Phi(i\nabla_x v(\cdot,x))\Omega_0 \rangle_\mathcal{F} = 0\,.
\end{equation}
As a consequence,
$(\nabla_xP_*)(x)P_*(x)= |\nabla_x\Omega(x)\rangle \langle\Omega(x)|$.
Hence we obtain for $1\leq j \leq N$
\begin{equation}\label{Rexp}
R(\delta,x)(\nabla_{x_j} P_*)(x)P_*(x) =i\, | U(x)\,R_{\rm f}(\delta)\,
\Phi(i\nabla_{x_j} v(\cdot,x))\Omega_0\rangle
\langle \Omega(x)|\,,
\end{equation}
where $R_{\rm f}(\delta) = (H_{\rm f} - i\delta)^{-1}$.
For (\ref{NGB1}) one therefore finds
\begin{eqnarray}
 \|R(\delta,x)(\nabla_{x_j} P_*)(x)P_*(x)\|^2_\mathcal{L(F)}&=&
 \|R_{\rm f}(\delta)\Phi(i\nabla_{x_j} v(x,\cdot))\Omega_0\|^2_\mathcal{F}\nonumber\\
&=& \left\| (|k| - i\delta)^{-1} k\,e^{ik\cdot x_j} \hat \rho_j(k)
|k|^{-\frac{3}{2}}\right\|^2_{L^2(\R^3)}  \label{RPest}
\end{eqnarray}
Whenever $\rho_j$ satisfies the infrared condition
$\hat \rho_j(k)|k|^{-\frac{3}{2}}\in L^2(\R^3)$,
 (\ref{RPest}) is bounded uniformly in $\delta$ since
 $\big|\,|k|/(|k|-i\delta)\,\big|\leq 1$.
In general we only assume that $\hat \rho_j(k)|k|^{-1}\in
L^2(\R^3)$. Using that $\hat \rho_j(k)$ is bounded uniformly according to Riemann-Lebesgue,
(\ref{RPest}) becomes
\begin{equation}\label{RPIRest}
 \left\| (|k| - i\delta)^{-1} k\,e^{ik\cdot x_j} \hat \rho_j(k)
|k|^{-\frac{3}{2}}\right\|^2_{L^2(\R^3)}
\leq
C_1\,\int_0^1\,d|k|\frac{|k|}{|k|^2+\delta^2} + C_2\leq C\ln{1/\delta}\,,
\end{equation}
for $\delta\in (0,\textstyle{\frac{1}{2}}]$.
To obtain the estimate (\ref{NGB2}), we differentiate (\ref{Rexp}) and find
\begin{eqnarray} \lefteqn{\hspace{-2cm}
\nabla_{x_i} \Big(R(\delta,x)(\nabla_{x_j} P_*)(x)P_*(x)\Big) =
i\, | U(x)\,R_{\rm f}(\delta)\Phi(i\nabla_{x_j} v(x,\cdot))\Omega_0\rangle
\langle\nabla_{x_i} \Omega(x)|}\nonumber
\\ &&
-\,\, | U(x)\,\Phi(i\nabla_{x_i} v(x,\cdot))\,R_{\rm f}
(\delta)\Phi(i\nabla_{x_j} v(x,\cdot))\Omega_0\rangle
\langle \Omega(x)|\nonumber\\
&& + \,\,i\, | U(x)\,R_{\rm f}(\delta)\Phi(i\nabla^{(2)}_{ij}
v(x,\cdot))\Omega_0\rangle
\langle \Omega(x)|\,.\label{63}
\end{eqnarray}
All three terms in (\ref{63}) can be bounded by using the same type of arguments
as in (\ref{RPest}) and (\ref{RPIRest}): For the first term use in addition that
$\|\nabla_{x_i} \Omega(x)\|_{\mathcal{H}_{\rm f}}<C$ and for the second term one has
to estimate the components in the $0$-boson sector and in the $2$-boson sector separately.
In summary we showed that  {\bf H}$_0^4$ (iii) is satisfied with
$\eta(\delta) = \delta\sqrt{\ln(1/\delta)}$,
and  with $\eta(\delta)=\delta$ if all charges satisfy the
infrared condition individually.

We are left to check for  Assumption {\bf h}$^4$.
Let $h^\epsi  = H_{\rm p}^\epsi\otimes {\bf 1}=
\sum_j h(-i\epsi\nabla_{x_j})\otimes {\bf 1}$, with  $h(p) = \sqrt{p^2 +1}$.
Essential self-adjointness of $h^\epsi + H_0$ on
$D(H_{\rm p}\otimes{\bf 1})\cap D(H_0)$
is a standard result, cf.\ Proposition 2.1 in \cite{Arai}.
We define $(Dh)^\epsi_j = (\nabla h)(-i\epsi\nabla_{x_j})\otimes {\bf 1}$, with
$\|(Dh)^\epsi\|_{\mathcal{L(H)}^{\oplus\,3N}}\leq 1$, and postpone the technical proof of the
following Lemma to the end of this section.

\begin{lem} \label{heLem}
$h^\epsi$ and $(Dh)^\epsi$ satisfy {\bf h}$^4$ (i) and (ii).
\end{lem}

We conclude that all  assumptions of Theorem \ref{NoGapTheorem} are
satisfied for the (SN) and the (IR) model and thus (\ref{HHdiagN}) holds.
However, (\ref{NTEquation}) follows from  (\ref{HHdiagN}) by the following Lemma
and an argument like (\ref{Gap1}).
\end{proof}

\begin{lem}
There is a constant $C<\infty$ such that for $\epsi>0$ sufficiently small
\[
\left\| \left(
H^\epsi_{\rm diag}\, -\,\mathcal{U}^*\,  H^\epsi_{\rm eff}\, \mathcal{U}\,
\right)\,P_*\,\right\|_\mathcal{L(H)}\leq \,\epsi^2\,C\,.
\]
\end{lem}

\begin{proof}
In order to apply {\bf h$^4$} (i) to $\mathcal{U}(x)$,   we have to extend
$\mathcal{U}(x):\mbox{Ran}P_*(x)\to \C$ defined in (\ref{UCDef}) to
a map $\widetilde{ \mathcal{U}}(\cdot)\in C^4_{\rm b}( \R^{3N},\mathcal{L(F)})$ first.
To this end let $\widetilde{\mathcal{U}}(x) =|\Omega_0\rangle \langle \Omega(x)|$
and note that $\widetilde{\mathcal{U}}^*\,\widetilde{\mathcal{U}}=P_*$.
With this definition one finds
\begin{eqnarray*}
H^\epsi_{\rm diag}\,P_* & = & H_0\,P_* + P_*\,h^\epsi\,P_* = E\,P_*  +   P_*\,
\widetilde{\mathcal{U}}^*\,h^\epsi\,\widetilde{\mathcal{U}}\,P_*
+   P_*\, [\,h^\epsi,\,
\widetilde{\mathcal{U}}^*\,]\,
\widetilde{\mathcal{U}}\,P_* \\
&=&
\mathcal{U}^*\,  H^\epsi_{\rm eff}\, \mathcal{U}\,P_*
+   P_*\, [\,h^\epsi,\,
\widetilde{\mathcal{U}}^*\,]\,
\widetilde{\mathcal{U}}\,P_*\,,
\end{eqnarray*}
and we are left to show that $\| P_*\, [\,h^\epsi,\,
\widetilde{\mathcal{U}}^*\,]\,
\widetilde{\mathcal{U}}\,P_*
\|=\mathcal{O}(\epsi^2)$.
Using {\bf h$^4$} (i) with $A=\widetilde{ \mathcal{U}}^*$  we find that
\[
 P_*\, [\,h^\epsi,\,
\widetilde{\mathcal{U}}^*\,]\,
\widetilde{\mathcal{U}}\,P_* =
-i\epsi \, P_*\, (\nabla_x\,\widetilde{\mathcal{U}}^*)\cdot
(Dh)^\epsi\,\widetilde{\mathcal{U}}\,P_* + \mathcal{O}(\epsi^2)\,.
\]
However, according to (\ref{Omeg})
\[
P_*(x)(\nabla_x\,\widetilde{\mathcal{U}}^*)(x) = |\Omega(x)\rangle\langle\Omega(x),\,
\nabla_x\,\Omega(x)\rangle\langle\Omega_0|= 0\,,
\]
and thus the desired result follows.
\end{proof}

\begin{proof}[Proof of Lemma \ref{heLem}]
Heuristically {\bf h}$^4$ (i) and (ii) hold, because they  are just special cases of
the  expansion of a commutator of pseudodifferential operators. However, since
$h^\epsi$ is unbounded and $A$ is only 4-times differentiable, we need to check the
estimates ``by hand''.

For notational simplicity we restrict ourselves to the case $N=1$, from
which the general case follows immediately.
Let $g(p) = 1/\sqrt{p^2+1}$, $g^\epsi=g(-i\epsi\nabla_x)\otimes{\bf 1}$
and $A\in C_{\rm b}^4(\R^3,\mathcal{H}_{\rm f})$,
then $|\cdot|^s \widehat g\in L^1(\R^3)$ for $s\in \{0,4\}$ and thus
for $\psi\in\mathcal{S}$
\begin{eqnarray}\lefteqn{\label{gA}
\Big( g^\epsi A \psi\Big)(x)
=
\int dy\,\widehat g(y)\,A(x-\epsi y)\,\psi(x-\epsi y)}\nonumber\\
&=&
\int dy\,\widehat g(y)\,\left( A(x) -\epsi y\cdot\nabla A(x) + \epsi^2 \int_0^1 ds\,
\langle y, \nabla^{(2)}A(x-s\epsi y)\,y\rangle\right)
\psi(x-\epsi y)\nonumber\\
&=&
\Big( A g^\epsi \psi\Big)(x) - i\,\epsi  \Big( \nabla A\cdot\nabla
g^\epsi \psi\Big)(x) \nonumber\\&&\hspace{2cm}+\,
\int dy\,\widehat g(y)\, \epsi^2 \int_0^1 ds\,
\langle y, \nabla^{(2)}A(x-s\epsi y)\,y\rangle
\psi(x-\epsi y)\,.
\end{eqnarray}
From (\ref{gA}) one concludes after a lengthy but straightforward computation involving
several integrations by parts that
\[
\epsi^2\Delta_x\,[g^\epsi,A] = - i\,\epsi  \nabla A\cdot(\nabla g)^\epsi\,(\epsi^2\Delta_x) + R
\]
with
\[
\|R\,\|\leq \,C\,\sum_{j=2}^4 \,\epsi^j \sup_{x\in \R^{3N},\,|\alpha|=j} \|\partial_x^\alpha
A(x)\|_{\mathcal{L}(\mathcal{H}_{\rm f})}\,.
\]
Hence we find
\[
[h^\epsi,A] = [(1-\epsi^2\Delta_x)g^\epsi,A] = (1-\epsi^2\Delta_x)[g^\epsi,A] -
[\epsi^2\Delta_x,A]g^\epsi
= -i\,\epsi  \nabla A\cdot (Dh)^\epsi + R'
\]
with
\[
\|R'\,\|\leq \,C'\,\sum_{j=2}^4 \,\epsi^j \sup_{x\in \R^{3N},\,|\alpha|=j} \|\partial_x^\alpha
A(x)\|_{\mathcal{L}(\mathcal{H}_{\rm f})}\,.
\]
This proves {\bf h}$^4$ (i). By the same type of arguments one shows also {\bf h}$^4$ (ii).
\end{proof}
\smallskip

\noindent {\bf Acknowledgments.}\quad
I am grateful to Herbert Spohn for suggesting the massless Nelson model as an application
for a space-adiabatic theorem without gap, as well as for numerous valuable discussions, remarks
and hints concerning the literature.
Parts of this work developed during a stay of the author at the Univerit\'e de Lille and
I thank Stephan De Bi\`evre and Laurent Bruneau for hospitality and
for a helpful introduction to Reference \cite{Arai}.
For critical remarks which lead to an improved presentation a thank Detlef D\"urr.

\end{document}